\documentclass[12pt,a4]{revtex4}
\usepackage[british]{babel}
\usepackage{amssymb,amsmath,amsthm, bbm, mathrsfs}
\usepackage[T1]{fontenc}
\usepackage{latexsym} 
\usepackage{times}
\usepackage{color}
\usepackage{xcolor}
\usepackage{graphicx}
\usepackage[T1]{fontenc}
\usepackage{comment}
\usepackage{t-angles}
\usepackage[framemethod=tikz]{mdframed}
\definecolor{refkey}{rgb}{0,0,1}
\definecolor{labelkey}{rgb}{1,0,0}
\usepackage{cleveref}
\colorlet{shadecolor}{gray!20}

\definecolor{ag}{rgb}{0.29, 0.33, 0.13}
\definecolor{darkblue}{rgb}{0.0, 0.0, 0.55}
\definecolor{darkcerulean}{rgb}{0.03, 0.27, 0.49}
\definecolor{darkpowderblue}{rgb}{0.0, 0.2, 0.6}
\definecolor{britishracinggreen}{rgb}{0.0, 0.26, 0.15}
\newenvironment{blu}{\color{darkpowderblue}}{}
\newenvironment{mgg}{\color{magenta}}{}

\newcommand{\bblu}{\begin{blu}}
\newcommand{\eblu}{\end{blu}}
\newcommand{\bmag}{\begin{mgg}}
\newcommand{\emag}{\end{mgg}}
\newcommand{\bas}{\begin{mgg}}
\newcommand{\eas}{\end{mgg}}
\definecolor{refkey}{rgb}{0,0,1}
\definecolor{labelkey}{rgb}{1,0,0}
\def\<{{\langle}} 
\def\>{{\rangle}}

\def\note#1{{}}

\def\note#1{} 
\def\beq{\begin{equation}} 
\def\eeq{\end{equation}}

\newcounter{zlist} 

\newcounter{blist} 
 
\newcounter{rlist} 
 
 \def\stac#1{\raise-.2cm\hbox{$\stackrel{\displaystyle\otimes}{\scriptscriptstyle{#1}}$}}

\def\cten#1{\raise-.2cm\hbox{$\stackrel{\displaystyle\widehat{\otimes}}{\scriptscriptstyle{#1}}$}}

\def\Label#1{\label{#1}\ifmmode\llap{[#1] }\else 
\marginpar{\smash{\hbox{\tiny [#1]}}}\fi} 
\def\Label{\label} 
\theoremstyle{definition}

\theoremstyle{remark}

\newcounter{c} 
 
\newcommand{\etyk}[1]{\vspace{-7.4mm}$$\begin{equation}\Label{#1} 
\addtocounter{c}{1}} 
\renewcommand{\]}{\ifnum \value{c}=1 $$\else \end{equation}\fi} 
\newcommand{\Cc}{\mathcal{C}}

\def\*C{{}^*\hspace*{-1pt}{\Cc}}
\def\text#1{{\rm {\rm #1}}}

\def\1{\mathbf{1}}

\newcounter{mnotecount}[section]
\renewcommand{\themnotecount}{\thesection.\arabic{mnotecount}}
\newcommand{\mnote}[1]%{}
{\protect{\stepcounter{mnotecount}}$^{\mbox{\footnotesize
$%\!\!\!\!\!\!\,
\bullet$\themnotecount}}$ \marginpar{%\color{red}%
\raggedright \tiny\em
$\bullet$\themnotecount: #1} }
\numberwithin{equation}{section}
\usepackage{pgfplots}
\pgfplotsset{width=8cm,compat=1.10}
\definecolor{dark_green}{rgb}{0.0, 0.5, 0.0}
\usepgfplotslibrary{fillbetween}
\begin{document} 
\vspace*{-2cm}
\title{Towards modified bimetric theories within non-product spectral geometry} 
\author{Arkadiusz Bochniak}
%\thanks{Authors acknowledge support by NCN grant OPUS 2016/21/B/ST1/02438}
\affiliation{Institute of Theoretical Physics, Jagiellonian University,
\hbox{prof.\ Stanis\l awa \L ojasiewicza 11, 30-348 Krak\'ow, Poland.}}
%%%%%%%%%%%%%%%%%%
%%\pacs{23.23.+x, 56.65.Dy}
%%%%%%%%%%%%%%%%%%%%%%%%%%%%%%%%%%%%%%%%%%%%%%%%%%%%

\begin{abstract} 
We discuss class of doubled geometry models with diagonal metrics. Based on the analysis of known examples we formulate a hypothesis that supports treating them as modified bimetric gravity theories. Certain steps towards the generic case are then performed.  
\end{abstract} 

\maketitle 
 \vspace*{-1cm}
%%%%%%%%%%%%%%%%%%%%%%%%%%%%%%%%%%%%%%%%%%%%%%%%%%%%  

%%%%%%%%%%%%%%%%%%%%%%%%%%%%%%%%%%%%%%%%%%%%%%%%%%%%  
\section{Introduction}

The description of gravity in terms of geometric objects is the cornerstone of Einstein's General Relativity and leads to an intriguing possibility of geometrizing all of the fundamental interactions. One of the existing proposal is based on the noncommutative geometry \cite{Co94} - a framework that puts on equal footing both the metric structure of manifolds, Yang-Mills-type theories and also the Higgs mechanism. The spectral description of manifolds \cite{Co93} can be generalized into other than classical geometries like discrete spaces and their products with manifolds. The latter one leads to the definition of the so-called almost-commutative geometries that were successfully applied to the description of gauge theories \cite{Co96, Co96a, ChaCoMa07}. Appropriate choice of the finite space allows e.g. for the formulation of the noncommutative Standard Model of Particle Physics. In this case the finite geometry is build on the matrix algebra $\mathbb{C}\oplus\mathbb{H}\oplus M_3(\mathbb{C})$ whose choice is dictated by the gauge group of the model \cite{vSbook}. 

Yet another model of this type, but much more simpler, is the one studied by Connes and Lott \cite{CoLo91}, where the finite algebra is just $\mathbb{C}\oplus \mathbb{C}$ and corresponds to the two points. In this case, the product space can be thought of as $M\times \mathbb{Z}_2$, that is, we have two copies of the same manifold. One can further generalize this geometry and can allow for two distinct metrics on these two sheets \cite{sitarz2019}. Such a doubled geometry is beyond the usual almost-commutative framework and therefore is of a non-product type. Since the spectral action principle applied to a single copy produces the Hilbert-Einstein action, the natural question of the form of an action functional for this non-product type of geometries arises. The answer for generic choices of metrics is not known yet, but in the case of the Friedmann-Lema{\^i}tre-Robertson-Walker (FRLW) type of Euclidean metrics this was done analitycally \cite{sitarz2019, BS2021}, and the stability of certain solutions was also analysed \cite{BS2021}. It was demonstrated therein that the interaction between the metrics resembles features characteristic to bimetric gravity models \cite{HaRo12,AKMS13}. Despite numerous similarities, certain significant differences are also present. In particular, the interaction potential for bimetric model is a polynomial one, while for the two-sheeted model it is a rational function. Further similarities and differences for generic metrics were recently analysed in \cite{BoSi22}, where yet another interpretation of this model in terms of interacting branes was proposed.  

In this note we discuss yet another class of models beyond the FLRW framework. We illustrate the generally claimed features on the simplified example - the so-called Hopf model. In this case the interaction potential has nontrivial logarithmic terms but it still possesses bimetric gravity characteristics. Finally, we make a general comment on the doubled geometry models which may allow for its future numerical studies.

\section{The generic diagonal model}

A framework of spectral geometry, allowing for an equivalent description of geometric objects in terms of algebraic data, originates from the observation that the geometry of a compact spin Riemannian manifold $M$ can be encoded in the collection of data $(C^\infty(M),L^2(M),D_M)$ \cite{Co93}, where $L^2(M)$ is the Hilbert space of square-integrable spinors, and $D_M$ is the Dirac operator, which can be written locally (with the use of the spin connection $\omega$) as $i\gamma^\mu(\partial_\mu +\omega_\mu)$.

This system of data is a prerequisite for the notion of a spectral triple, a set $(A,H,D)$ consisting of a unital $\ast$-algebra $A$ represented in a faithful way on a Hilbert space $H$, on which the (possibly unbounded) densely defined self-adjoint operator $D$ acts.  In the generic case it is assumed that the commutators $[D,a]$, $a\in A$, are well-defined and can be (uniquely) extended to an element from $B(H)$, bounded operators on $H$. Furthermore, the resolvent of $D$ has to be compact. Several further comptability conditions are imposed for certain applications \cite{Li18, vSbook}. 

In addition to the aforementioned canonical spectral triple associated to a manifold $M$, the finite dimensional ones are well-understood \cite{Kra98, PaSi98}. In this case both an algebra $A_F$ and a Hilbert space $H_F$ are finite dimensional, and $D_F$ is just a matrix. One can go one step further and consider products of spectral triples considered so far. The almost-commutative geometry is a result of such a construction, where the first spectral triple in the product is the canonical one for a manifold $M$, and the other one is finite. In the case with $\dim M=4$, the Dirac operator for the resulting triple is (pointwisely) $D_M\otimes 1+ \gamma_5\otimes D_F$, where $\gamma_5$ is the usual grading in the Clifford algebra associated to the manifold $M$. 

However, even for the product space with finite part being just the two points set, this is not the most general Dirac operator one can consider. Indeed, the operator 
\begin{equation}
\mathcal{D}=\begin{pmatrix}
D_1 & \gamma\Phi\\
\gamma\Phi^\ast & D_2
\end{pmatrix}
\end{equation}
with a field $\Phi$, which for our purposes is taken to be a constant, is an example of another candidate \cite{sitarz2019}. Here $D_1,D_2$ are two Dirac operators for $M$, but considered with two different Riemannian metrics $g_1, g_2$. The operator $\gamma$ is a straightforward generalization of $\gamma_5$: $\gamma^\ast=\gamma$, $\{\gamma,\gamma^a\}=0$ for all anti-Hermitian $\gamma^a$ generating the Clifford algebra, $\{\gamma^a,\gamma^b\}=-2\delta^{ab}1$, but now $\gamma^2=\kappa=\pm 1$ (instead of requiring $\kappa=1$). These models are refered to as the doubled geometries \cite{BS2021}.

We consider geometries of this type with the metric on each sheet chosen to be of the form
\begin{equation}
\label{metryka}
ds^2=  \sum\limits_{j=0}^3 a_j^2 \left( dx^j \right)^2,
\end{equation}
where $a_j$, for $j=0,1,2,3$, are constants. The spin connection $\omega$ is identically zero since for the coframe $\{\theta^a\}$ we have $d\theta^a =0$ for every $a=0,\ldots,3$, and the resulting Dirac operator therefore reads, $D= \sum\limits_{j=0}^3 \frac{1}{a_j}\gamma^j\partial_j$. The corresponding doubled geometry constructed out of these two sheets is therefore described by a Dirac operator of the form
\begin{equation}
\mathcal{D}= \sum\limits_{j=0}^3 A_j \gamma^j \partial_j + \gamma F,
\end{equation}
where
\begin{equation}
A_j=\begin{pmatrix}
\frac{1}{a_{1,j}} & \\
& \frac{1}{a_{2,j}}
\end{pmatrix},
\hspace{10pt}
F=
\begin{pmatrix}
 & \Phi \\
 \Phi^\ast &
\end{pmatrix}.
\end{equation}

The associated Laplace operator is hence given by
\begin{equation}
\mathcal{D}^2=-\sum\limits_{j=0}^3 A_j^2 \partial_j^2 
+ \sum\limits_{j=0}^3[F,A_j]\gamma\gamma^j \partial_j 
+ \kappa F^2,
\end{equation}
and one can then easily read the decomposition of its symbol into the homogeneous parts, $\sigma_{\mathcal{D}^2}=\mathfrak{a}_0+\mathfrak{a}_1+\mathfrak{a}_2$.

Since our first goal is to determine the leading terms of the spectral action,
\begin{equation}
\label{spectral_action}
\begin{aligned}
\mathcal{S}(\mathcal{D})&=\Lambda^4 \, \mathrm{Wres}(\mathcal{D}^{-4})+c\Lambda^2 \, \mathrm{Wres}(\mathcal{D}^{-2})\\
&=\int_M\int_{\|\xi\|=1} 
\left(\Lambda^4\, \mathrm{Tr}\, \mathrm{Tr}_{Cl} \, {\mathfrak b}_0^2+c \Lambda^2 \, \mathrm{Tr}\, \mathrm{Tr}_{Cl} {\mathfrak b}_2 \right),
\end{aligned}
\end{equation}
we have to find the symbol of the inverse of the Laplace operator, $\sigma_{\mathcal{D}^{-2}}=\mathfrak{b}_0+\mathfrak{b}_1+\mathfrak{b}_2+...$, what can be achieved by using the standard methods of pseudodifferential calculus \cite{gilkey}. (In the above equation $\mathrm{Tr}_{Cl}$ denotes the trace performed over the Clifford algebra and $\mathrm{Tr}$ is the usual matrix trace over two-by-two matrices.)

In our case we get
\begin{equation}
\mathfrak{b}_0= \left(  \sum\limits_{j=0}^3 A_j^2 \xi_j^2\right)^{-1}, \qquad \mathfrak{b}_2=\mathfrak{b}_0\mathfrak{a}_1\mathfrak{b}_0\mathfrak{a}_1\mathfrak{b}_0-\mathfrak{b}_0\mathfrak{a}_0\mathfrak{b}_0,
\end{equation}
so that
\begin{equation}
\mathrm{Tr}_{Cl}(\mathfrak{b}_2)=-4\kappa \mathfrak{b}_0\left(\sum\limits_{j=0}^3 [F,A_j]\mathfrak{b}_0[F,A_j]\xi_j^2 + F^2\right)\mathfrak{b}_0.
\end{equation}

The only nonzero elements of the matrix $\mathfrak{b}_0$ are on its diagonal and they are equal to
\begin{equation}
(\mathfrak{b}_0)^i_{\ i}= 
\frac{1}{ \sum\limits_{j=0}^3 A_{i,j}^2 \xi_j^2},
\end{equation}
where $A_{i,j}\equiv (A_j)^i_{\ i}=\frac{1}{a_{i,j}}$, and as a result of a straightforward computation we get
\begin{equation}
\begin{aligned}
&\mathrm{Tr}\left(-4\kappa \mathfrak{b}_0\sum\limits_{j=0}^3 [F,A_j]\mathfrak{b}_0[F,A_j]\mathfrak{b}_0\xi_j^2 \right)\\
&= 
4 \kappa|\Phi|^2 \sum\limits_{j,k=0}^3
\frac{ (A_{2,j}-A_{1,j})^2 (A_{1,k}^2+A_{2,k}^2) }
{\left( \sum\limits_{l=0}^3 A_{1,l}^2 \xi_l^2 \right)^2
\left( \sum\limits_{l=0}^3 A_{2,l}^2 \xi_l^2 \right)^2} \xi_j^2 \xi_k^2.
\end{aligned}
\end{equation}
The resulting spectral action is therefore of the form
\begin{equation}
\mathcal{S}(\mathcal{D})\sim \int_{M}\left(\Lambda_e^2 \mathcal{S}_{\Lambda_e} +\alpha\widehat{V}(g_1,g_2)\right),
\end{equation}
with
\begin{equation}
\mathcal{S}_{\Lambda_e}=\int_{\|\xi\|=1}\left\{\left( \sum\limits_{j=0}^3 A_{1,j}^2 \xi_j^2 \right)^{-2}+ \left( \sum\limits_{j=0}^3 A_{2,j}^2 \xi_j^2 \right)^{-2}\right\}
\end{equation}
and
\begin{equation}
\label{eq:pot}
\widehat{V}(g_1,g_2)=\sum\limits_{j,k=0}^3 (A_{2,j}-A_{1,j})^2 (A_{1,k}^2+A_{2,k}^2)
\int_{\|\xi\|=1}\frac{  \xi_j^2 \xi_k^2 }
{\left( \sum\limits_{l=0}^3 A_{1,l}^2 \xi_l^2 \right)^2
	\left( \sum\limits_{l=0}^3 A_{2,l}^2 \xi_l^2 \right)^2}
\end{equation}
where we have already introduced effective parametrization,
\begin{equation}
\Lambda_e^2 = \frac{12}{c}(\Lambda^2 - c\kappa |\Phi|^2), \qquad \alpha=12|\Phi|^2\kappa,
\end{equation}
and ommited the irrelevant global multiplicative constant. Therefore, the problem of finding the potential term describing the interaction between the two diagonal metrics reduces to compute linear combination of the integrals of the form
\begin{equation}
\int_{\|\xi\|=1} 
 \frac{ \xi_j^2 \xi_k^2 }
{\left( \sum\limits_{l=0}^3 A_{1,l}^2 \xi_l^2 \right)^2
	\left( \sum\limits_{l=0}^3 A_{2,l}^2 \xi_l^2 \right)^2}.	
\end{equation}
Moreover, from the Eqn. \eqref{eq:pot} it immediately follows that $\widehat{V}(g_1,g_2)=\widehat{V}(g_2,g_1)$, that is, $\widehat{V}$ is symmetric under the interchange $g_1\leftrightarrow g_2$. We further conjecture that the potential term can be written as
\begin{equation}
\widehat{V}(g_1,g_2)=2\pi^2\mathbb{V}\left(\sqrt{g_2^{-1}g_1}\right)\sqrt{\det\, g_2}
\end{equation}
for some function $\mathbb{V}$. By the symmetry of $\widehat{V}$, to prove this claim it is enough to show that the function $\mathbb{V}'(g_1,g_2):=\frac{\widehat{V}(g_1,g_2)}{2\pi^2 \sqrt{\det g_2}}$ depends only on the eigenvalues of $\sqrt{g_2^{-1}g_1}$. We illustrate this hypothesis on a simple nontrivial example - the Hopf model - discussed in the forthcoming section.

\section{The Hopf model}
We consider here models with diagonal metrics given by $g_{00}=g_{11}=b^2$ and $g_{22}=g_{33}=a^2$, for which then have
\begin{equation}
(\mathfrak{b}_0)^1_{\ 1} = \frac{a_1^2b_1^2}{a_1^2 (\xi_0^2 +\xi_1^2) +b_1 (\xi_2^2+\xi_3^2)}, \qquad (\mathfrak{b}_0)^2_{\ 2} = \frac{a_2^2b_2^2}{a_2^2 (\xi_0^2 +\xi_1^2) +b_2 (\xi_2^2+\xi_3^2)}
\end{equation}
and
\begin{equation}
\begin{aligned}
\widehat{V}(g_1,g_2)&=\frac{(b_1-b_2)^2}{b_1^2b_1^2}\int_{\|\xi\|=1}(\xi_0^2+\xi_1^2)\det(\mathfrak{b}_0)\mathrm{Tr}(\mathfrak{b}_0)\\
&+\frac{(a_1-a_2)^2}{a_1^2a_1^2}\int_{\|\xi\|=1}(\xi_2^2+\xi_3^2)\det(\mathfrak{b}_0)\mathrm{Tr}(\mathfrak{b}_0).
\end{aligned}
\end{equation}

In order to parametrize the three-sphere $\|\xi\|=1$ we use here the following Hopf-like coordinates which resemble the symmetry of the system:
\begin{equation}
\xi_0=\cos\theta\cos\varphi, \quad \xi_1=\cos\theta\sin\varphi, \quad \xi_2=\sin\theta \cos\psi, \quad \xi_3=\sin \theta \sin\psi.
\end{equation}
The angle $\theta$ is taken from $\left[0,\frac{\pi}{2}\right]$, while $0\leq \psi \leq 2\pi$, and the surface element in these coordinates is then given by $dS=\cos\theta\sin\theta  d\theta d\varphi d\psi$.

As a result, we get
\begin{equation}
\begin{aligned}
&\det(\mathfrak{b}_0)\mathrm{Tr}(\mathfrak{b}_0) \\
=&\frac{a_1^2a_2^2b_1^2b_2^2\left[a_1^2b_1^2(a_2^2\cos^2\theta +b_2^2\sin^2\theta)+a_2^2b_2^2(a_1^2 \cos^2\theta +b_1^2\sin^2\theta)\right]}{\left[a_1^2 \cos^2\varphi(a_1^2\cos^2\theta +b_1^2\sin^2\theta)\sin^2\varphi\right]^2\left[a_2^2 \cos^2\varphi(a_2^2\cos^2\theta +b_2^2\sin^2\theta)\sin^2\varphi\right]^2}.
\end{aligned}
\end{equation}
Let us introduce the following notation
\begin{equation}
I_{\mu,c}=\int_{\|\xi\|=1}\frac{\xi_\mu^2\cos^2\theta \, dS}{\left[a_1^2 \cos^2\varphi(a_1^2\cos^2\theta +b_1^2\sin^2\theta)\sin^2\varphi\right]^2\left[a_2^2 \cos^2\varphi(a_2^2\cos^2\theta +b_2^2\sin^2\theta)\sin^2\varphi\right]^2},
\end{equation}
and
\begin{equation}
I_{\mu,s}=\int_{\|\xi\|=1}\frac{\xi_\mu^2\sin^2\theta \, dS}{\left[a_1^2 \cos^2\varphi(a_1^2\cos^2\theta +b_1^2\sin^2\theta)\sin^2\varphi\right]^2\left[a_2^2 \cos^2\varphi(a_2^2\cos^2\theta +b_2^2\sin^2\theta)\sin^2\varphi\right]^2},
\end{equation}
and notice that 
\begin{equation}
I_{0,c}=I_{1,c}=I_{2,c}=I_{3,c}, \qquad I_{0,s}=I_{1,s}, \qquad I_{2,s}=I_{3,s},
\end{equation}
so that
\begin{equation}
\begin{aligned}
\widehat{V}(g_1,g_2)&=2a_1^2a_2^2b_1^2b_2^2 \left\{a_1^2a_2^2(b_1^2+b_2^2)\left[\frac{(b_1-b_2)^2}{b_1^2b_2^2}I_{0,c}+\frac{(a_1-a_2)^2}{a_1^2a_2^2}I_{0,s}\right]\right. \\
&\left.+b_1^2b_2^2(a_1^2+a_2^2)\left[\frac{(b_1-b_2)^2}{b_1^2b_2^2}I_{0,s}+\frac{(a_1-a_2)^2}{a_1^2a_2^2}I_{2,s}\right]\right\}.
\end{aligned}
\end{equation}

In order to find the final form of the potential it remains to compute the integrals $I_{0,c}$, $I_{0,s}$ and $I_{2,s}$. The result reads,
\begin{equation}
\widehat{V}(g_1,g_2)=\frac{2\pi^2}{(a_2b_1-a_1b_2)(a_2b_1+a_1b_2)^2}\left(F(a_1,a_2,b_1,b_2)+G(a_1,a_2,b_1,b_2)\right),
\end{equation}
where
\begin{equation}
F(a_1,a_2,b_1,b_2)=4a_1^2a_2^2b_1^2b_2^2(a_1-a_2)(b_1-b_2)\log\left(\frac{a_1 b_2}{a_2b_1}\right),
\end{equation}
and
\begin{equation}
\begin{aligned}
G(a_1,a_2,b_1,b_2)=(a_2^2b_1^2-a_1^2b_2^2)\left[a_1^2 b_1^2 a_2 (b_1-2b_2)\right. & \left. +a_2^2b_2^2 a_1(b_2-2b_1)\right. \\ &+\left. a_1^3 b_1^2 b_2 + a_2^3 b_2^2 b_1\right].
\end{aligned}
\end{equation}

We observe that for $a_1=a_2$ the potential reduces to $\widehat{V}(g_1,g_2)=2\pi^2 a_1^2 (b_1-b_2)^2$, while for $b_1=b_2$ we have $\widehat{V}(g_1,g_2)=2\pi^2 (a_1-a_2)^2b_1^2$.

Since the logarithm vanishes if and only if $\frac{a_1}{a_2}=\frac{b_1}{b_2}$ it would be, in principle, interesting to consider the limit of $\widehat{V}(g_1,g_2)$ when $b_1$ tends to $b_2 \frac{a_1}{a_2}$. The value of the function $V$ is indeterminated in this case, but the limit may still exists. Indeed, as a result we get
\begin{equation}
\lim_{b_1\rightarrow b_2\frac{a_1}{a_2}} \widehat{V}(g_1,g_2)=\frac{b_2^2}{a_2^2}(a_1-a_2)^2(a_1^2+a_2^2).
\end{equation}   

Introducing the new variable $x=\frac{b_1}{b_2}$ and $y=\frac{a_1}{a_2}$ we can write
\begin{equation}
\widehat{V}(g_1,g_2)=2\pi^2 \mathbb{V}\left(\sqrt{g_{2}^{-1}g_1}\right)\sqrt{\det\, g_2}, 
\end{equation}
where

\begin{equation}
\mathbb{V}\left(\sqrt{g_{2}^{-1}g_1}\right)=\frac{4x^2y^2(x-1)(y-1)}{(x-y)(x+y)^2}\log\left(\frac{y}{x}\right)+x^2y^2+1-2xy\frac{xy+1}{x+y}.
\end{equation}
We observe that 
\begin{equation}
\mathbb{V}\left(\sqrt{g_2^{-1}g_1}\right)\sqrt{\det\, g_2}=\mathbb{V}\left(\sqrt{g_1^{-1}g_2}\right)\sqrt{\det\, g_1},
\end{equation}
what illustrates the hypothesis.
%%%%%%%%%%%%%%%%%%%%%%%%%
%%%%%%%%%%%%%%%%%%%%%%%%%
\section{Comment on generic metrics}

The so far examined examples of doubled geometries suggest that these models can be thought of as certain modifications of bimetric theories as the potential term possesses features characteristic to this type of modified gravity theories. Despite the fact that series of non-trivial examples are already analysed, the derivation of the action in the generic case is still an open problem. In the approach we are using the main chalenge is related with the computation of certain integrals of rational functions defined over higher spheres: 
\begin{equation}
I=\int_{\|\xi\|=1}\frac{d^4\xi}{A_{\mu\nu}\xi^\mu \xi^\nu},
\end{equation} 
with smooth $A_{\mu\nu}$, which can be further written as $A_{\mu\nu}=\Omega(\delta_{\mu\nu}+\epsilon_{\mu\nu})$ with $\Omega\in\mathbb{R}$ and $\epsilon_{\mu\nu}$ being symmetric and traceless. In the formula above $\xi^\alpha$ is the $\alpha$th coordinate of vector $\xi$.

We make here some comments on the analysis of doubled geometry models in case when the tensors $A_{\mu\nu}$ does not differ sufficiently from the diagonal ones. This is not identical to the situation where the metrics are small perturbation of the Euclidean ones - for the discussion of the latter we refer to \cite{BoSi22}.

Assuming that $\max\limits_{\mu,\nu}\|\epsilon_{\mu\nu}\|_{\infty}$ is sufficiently small we can expand in these parameters and write
\begin{equation}
I=\frac{1}{\Omega}\sum\limits_{m\ge 0}(-2)^m \epsilon_{\alpha_1\beta_1}\ldots \epsilon_{\alpha_m\beta_m}I^{\alpha_1\beta_1\ldots \alpha_m\beta_m}_{4,m},
\end{equation}
where
\begin{equation}
I^{\alpha_1\beta_1\ldots \alpha_m\beta_m}_{n,m}=\int_{\mathbb{S}^{n-1}}d^{n}\xi \xi^{\alpha_1}\xi^{\beta_1}\ldots \xi^{\alpha_m}\xi^{\beta_m}
\end{equation}
are polynomial integrals over higher spheres which can be evaluated generalizing the methods from \cite{othmani} - see also \cite{BoSi22} for further discussion. 

Denoting 
\begin{equation}
\gamma_j=\begin{cases}\alpha_k, \quad j=2k-1,\\
\beta_k, \quad j=2k \end{cases}
\end{equation}
we define $I^{\gamma_1\ldots \gamma_{2m}}=I^{\alpha_1\beta_1 \ldots \alpha_m\beta_m}_{4,m}$, and
let $\Delta^{\gamma_1...\gamma_{2m}}$ be the sum of product of deltas in $I^{\gamma_1...\gamma_{2m}}$, i.e.
\begin{equation}
I^{\gamma_1...\gamma_{2m}}=c_m \Delta^{\gamma_1...\gamma_{2m}}
\end{equation}
with some real number $c_m$, and $\Delta^{...}=\sum\delta^{..}...\delta^{..}$
Since $\epsilon$ is traceless not all terms in 
\begin{equation}\epsilon_{\gamma_1\gamma_2}...\epsilon_{\gamma_{2m-1}\gamma_{2m}}I^{\gamma_1...\gamma_{2m}}
\end{equation}
are nonzero. Let $\mathcal{N}_{2m}$ be the number of nonzero terms, and consider the following problem. Suppose the numbers $1,...,2m$ are given, and we would like to use them to fill in an $1\times 2m$ array $T$, with a given subdivision into $1\times 2$ subarrays $T=T_1|T_2|...|T_m$, as follows:
\begin{itemize}
\item In the first entry of $T_1$ we put $1$,
\item For every $j=1,...,m$, we have $T_j=\begin{bmatrix}a_j | b_j \end{bmatrix} $ with $a_j<b_j$,
\item For every $j=1,...,m$, $a_j<a_{j+1}$.
\end{itemize}
Then $\mathcal{N}_{2m}$ is a number of such fillings for which there is no $j$ such that $T_j$ is of the form $\begin{bmatrix} 2l-1 | 2l \end{bmatrix}$, for some $l=1,...,m$.

Since, by symmetry of $\epsilon$, any nonzero term in $\epsilon_{\gamma_1\gamma_2}...\epsilon_{\gamma_{2m-1}\gamma_{2m}}I^{\gamma_1...\gamma_{2m}}$ produces $\mathrm{tr}(\epsilon^m)$, we have
\begin{equation}
\epsilon_{\gamma_1\gamma_2}...\epsilon_{\gamma_{2m-1}\gamma_{2m}}I^{\gamma_1...\gamma_{2m}}=\mathcal{N}_{2m}c_m\mathrm{tr}(\epsilon^m),
\end{equation}
and the problem reduces to finding coefficients $c_m$. Since $\mathrm{area}(\mathbb{S}^3)=2\pi^2$ we get $c_1=\frac{\pi^2}{2}$. Moreover, by using the generalization  of \cite[Prop.~2]{othmani} (see also \cite[Prop.~A.2]{BoSi22} ) one can easily find the recursive formula for $c_m$:
\begin{equation}
c_m=\frac{c_{m-1}}{4+2(m-1)},
\end{equation}
and its solution reads 
\begin{equation}
c_m=\frac{4\pi^2}{(2m+2)!!}.
\end{equation}

As a result,
\begin{equation}
I=\frac{1}{\Omega}\left[2\pi^2 +\frac{2\pi^2}{3}\mathrm{tr}(\epsilon^2)+4\pi^2\sum_{m\ge 3}\frac{(-2)^m}{(2m+2)!!}\mathcal{N}_{2m}\mathrm{tr}(\epsilon^m)\right].
\end{equation}
In order to apply this result to a specific term of the action one has to first solve the combinatorial problem of finding the coefficients $\mathcal{N}_{2m}$, up to required order in $m$.  We postpone for the future research the problem of determining set of metrics for which the rate of convergence of the above series is satisfactory for all the terms that appear in the action functional.
 
%%%%%%%%%%%%%%%%%%%%%%%%%
%%%%%%%%%%%%%%%%%%%%%%%%%
\section{Conclusions and outlook}

The discussed doubled geometry model is an interesting possibility of going beyond the General Relativity. The explicit functional form of its action is derivable in the same way as the Hilbert-Einstein's one but with the use of a different geometry instead of the classical manifold. Here we extended the existing family of known examples for which the features characteristic to bimetric gravity models are present. We also made further steps towards the analysis of models that are beyond the class of such whose action is analytically computable. We remark that yet another approach based on a different type of noncommutativity can produce bimetric type of models \cite{Ce18}. It will be interesting to find some deeper relations between these two formulations - we postpone this for a future research.

%%%%%%%%%%%%%%%%%%%%%%

{\bf Acknowledgments}\\ The author thanks A.~Sitarz for helpful discussions and comments. The author acknowledges the support from the National Science Centre, Poland, Grant No. 2018/31/N/ST2/00701.

%%%%%%%%%%%%%%%%%%%%%%%%%
%%%%%%%%%%%%%%%%%%%%%%%%%

\noindent\rule{\textwidth}{1pt}

\end{document}